\documentclass[journal]{IEEEtran}

\usepackage{cite}
\usepackage{algorithm}
\usepackage[noend]{algpseudocode}
\usepackage{amsmath,amssymb,amsfonts}
\usepackage[pdftex]{graphicx}
\usepackage{subcaption}
\usepackage{textcomp}
\usepackage{xcolor}
\usepackage{pifont}
\usepackage{multirow}
\usepackage{makecell}
\usepackage{array}
\usepackage{diagbox}
\usepackage{CJKutf8}
\usepackage{soul}
\usepackage{hyperref}

\begin{document}

\title{Large Model Empowered Streaming Speech Semantic Communications}

\author{Zhenzi Weng,~\IEEEmembership{Member,~IEEE}, Zhijin Qin,~\IEEEmembership{Senior Member,~IEEE}, and Geoffrey Ye Li,~\IEEEmembership{Fellow,~IEEE}
\thanks{Zhenzi Weng and Geoffrey Ye Li are with the Department of Electrical and Electronic Engineering, Imperial College London, London SW7 2AZ, U.K. (e-mail: z.weng@imperial.ac.uk; geoffrey.li@imperial.ac.uk)(Corresponding author: Geoffrey Ye Li).}
\thanks{Zhijin Qin is with the Department of Electronic Engineering, Tsinghua University, Beijing 100084, China. She is also with the State Key Laboratory of Space Network and Communications and the Beijing National Research Center for Information Science and Technology, Beijing 100084, China (e-mail: qinzhijin@tsinghua.edu.cn).}
}

\maketitle
\begin{abstract}
In this paper, we introduce a large model-empowered streaming semantic communication system for speech transmission across various languages, named LSSC-ST. Specifically, we devise an edge-device collaborative semantic communication architecture by offloading the intricate semantic extraction and channel coding modules to edge servers, thereby reducing the computational burden on local devices. To support multilingual speech transmission, pre-trained large speech models are utilized to learn unified semantic features from speech in different languages, breaking the constraint of a single input language and enhancing the practicality of the LSSC-ST. Moreover, the input speech is sequentially streamed into the developed system as short speech segments, which enables low transmission latency without degrading the quality of the produced speech. A novel dynamic speech segmentation algorithm is proposed to further reduce the transmission latency by adaptively adjusting the duration of speech segments. According to simulation results, the LSSC-ST provides more accurate speech transmission and achieves a streaming manner with lower latency compared to the existing non-streaming semantic communication systems.

\end{abstract}

\IEEEpeerreviewmaketitle

\section{Introduction}
\IEEEPARstart{S}{emantic} communications have been proved to undergo unprecedented advancements over the past few years due to the booming of artificial intelligence (AI). To contend with the explosive growth of data traffic, deep learning (DL)-enabled semantic communications have been considered a promising solution to provide intelligent data transmission and address numerous bottlenecks in conventional communications~\cite{9770094,10639525}.

According to Shannon and Weaver~\cite{weaver1953recent}, communication can be categorized into three levels, including syntax communications, semantic communications, and pragmatic communications. The conventional communication paradigm falls under syntax communications, quantifies information at the bit level, and aims to achieve a low bit-error rate (BER) or symbol-error rate (SER). This bit-oriented communication framework ignores the meaning behind the transmission data, running counter to the ultimate goal of semantic exchange in wireless communications. In this context, semantic information has been investigated from different theoretical perspectives~\cite{floridi2004outline,6004632}. For the sake of efficient semantic representation, DL techniques have shown their potential to extract semantic information inherent in the source by leveraging the superior learning and fitting capabilities of sophisticated neural networks (NNs). DL-enabled semantic communications have attracted substantial attention and overcome the challenge on approximate semantic representation~\cite{8663966}.

DL-enabled semantic communications explore two transmission goals: source reconstruction and task execution. Particularly, Xie~\emph{et al.}~\cite{9398576} pioneered text semantic commutation system, DeepSC, by jointly designing the semantic and channel coding. Jiang~\emph{et al.}~\cite{9791409} devised a hybrid automatic repeat request (HARQ)-based semantic communication system to strengthen the transmission reliability of semantic information. Inspired by the flow of intelligence, Dong~\emph{et al.}~\cite{9954279} carried out a semantic communication system for image restoration, which enhances model flexibility across diverse transmission scenarios. Additionally, Xie~\emph{et al.}~\cite{9830752} developed task-oriented semantic communications for visual question-answering by fusing textual and visual semantic features at the receiver to infer the context. In~\cite{10431795}, Zhang~\emph{et al.} built a unified semantic communication framework for multitask execution by invoking a lightweight feature selection network.

In semantic communications for speech transmission, Weng~\emph{et al.}~\cite{9450827} proposed the first semantic communication system for speech reconstruction, named DeepSC-S. In~\cite{10038754}, Weng~\emph{et al.} further studied intelligent speech tasks in semantic communications, such as speech recognition and speech-to-text translation. Although task-oriented semantic communications for speech transmission have demonstrated superiority in serving AI tasks compared to the conventional speech transmission protocols, we are still facing the following challenges:
\begin{enumerate}
    \item[\emph{C1}:] \emph{The computational resources of user devices are insufficient for complicated feature extraction.}
    \item[\emph{C2}:] \emph{Existing works only support speech in one language and lack the adaptability for fine-tuning to others.}
    \item[\emph{C3}:] \emph{Significant transmission delay is caused by the requirement for the complete duration of input speech.}
\end{enumerate}

Large models have been applied to address various challenges in wireless communications~\cite{sheng2024beam}. This paper proposes a novel task-oriented large model-empowered semantic communication system for speech transmission, LSSC-ST, to address these challenges. It considers multilingual speech translation tasks, facilitating seamless speech communication across linguistic boundaries and reducing transmission latency by processing the input speech in a streaming manner. The main contributions of this paper can be summarized as follows:
\begin{itemize}
\item An edge-device collaborative semantic communication system for end-to-end speech translation is established, deploying large speech models on edge servers to perform semantic extraction from multilingual input speech and interacting with local devices via a reliable channel.

\item To avoid the delay attributed to waiting for the entire input speech, we devise an efficient mechanism that concurrently reads the next short speech segment and performs speech feature extraction on the current speech segment, achieving accurate speech translation with low communication latency.

\item To further improve the fluency of the translated speech, a dynamic speech segmentation algorithm is introduced to determine the duration of the current speech segment according to the amount of semantic information within the previous speech segment, which mitigates the discontinuity between adjacent translated speech segments.

\end{itemize}

The rest of this paper is structured as follows. In Section~\uppercase\expandafter{\romannumeral2}, the model of the large model-empowered streaming semantic communications is provided. Section~\uppercase\expandafter{\romannumeral3} presents the details of the proposed LSSC-ST. Section~\uppercase\expandafter{\romannumeral4} presents experimental results and Section~\uppercase\expandafter{\romannumeral5} concludes this paper.


\section{System Model}
This section briefly illustrates the considered system model for end-to-end speech translation across multiple languages. Then, the adopted performance metrics are introduced.

\subsection{Edge-Device Collaborative Communication Framework}
\begin{figure}[tbp]
\includegraphics[width=0.487\textwidth]{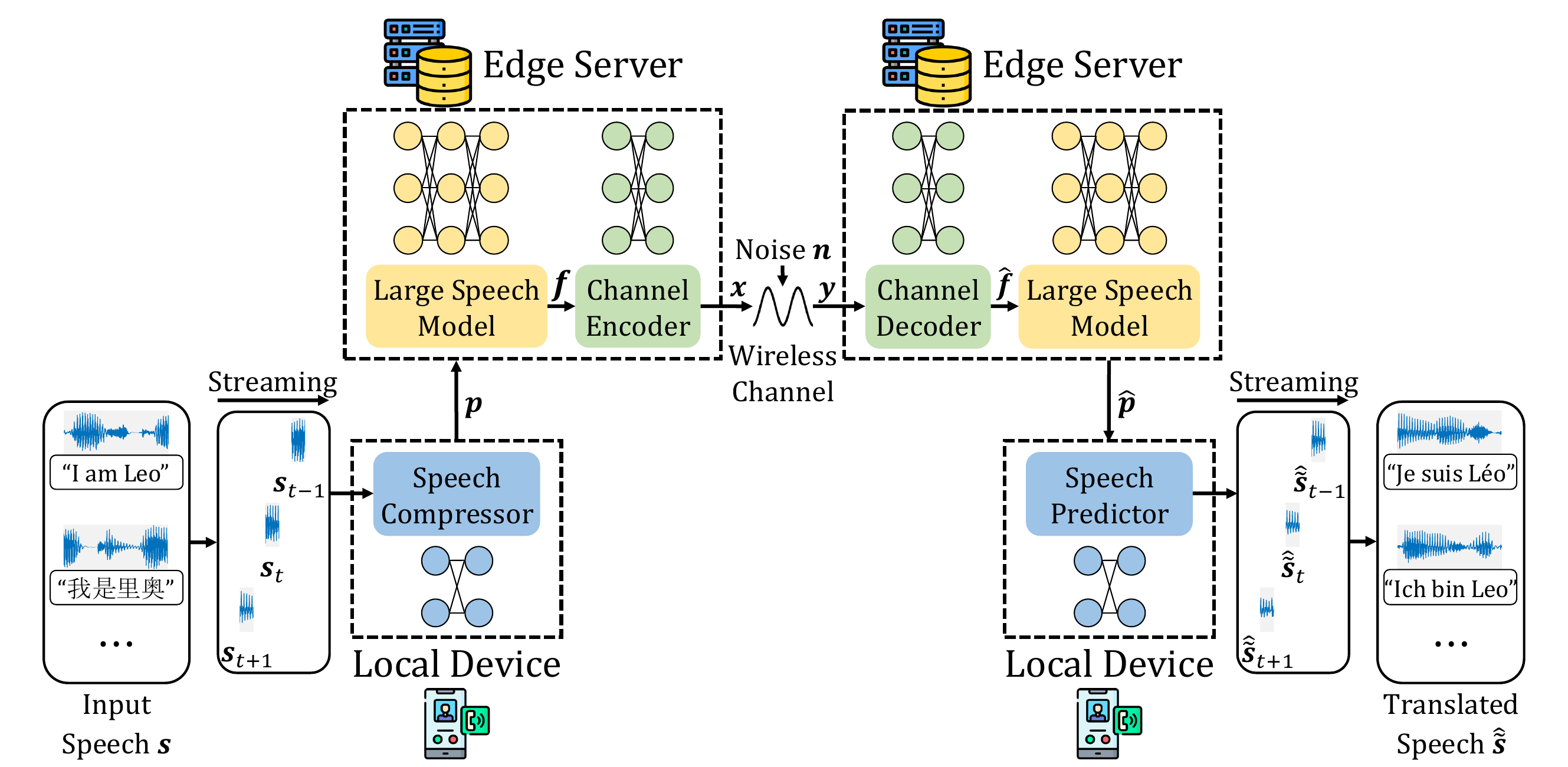}
\centering 
\caption{Model structure of large speech model-empowered streaming semantic communications for speech translation.}
\label{system model}
\end{figure}

The motivation of this work is to support the real-time speech translation task in semantic communications when users have various linguistic backgrounds. The model structure of the designed system is shown in Fig.~\ref{system model}. From the figure, the system input consists of speech in one of the supported languages, $\boldsymbol s=\left[{\boldsymbol s}_1,\;{\boldsymbol s}_2,\;...,\;{\boldsymbol s}_T\right]$, where ${\boldsymbol s}_t$ is $t$-th short speech segment in $\boldsymbol s$. ${\boldsymbol s}_t$ is fed into the speech compressor on the local device in a streaming manner to obtain the intermediate representation, $\boldsymbol p$, through a lightweight NN. It is worth mentioning that the following speech segment, ${\boldsymbol s}_{t+1}$, is being captured while the speech compressor is processing ${\boldsymbol s}_t$, and the data size of $\boldsymbol p$ is significantly smaller than that of ${\boldsymbol s}_t$ to streamline the interaction between the local device and the edge server, which addresses the preceding challenge,~\emph{C3}. Denote the speech compressor as ${\mathfrak T}_{\mathrm{SC}}(\cdot)$, then $\boldsymbol p$ is written as
\begin{equation}
\boldsymbol p={\mathfrak T}_{\mathrm{SC}}({\boldsymbol s}_t)\;\;\;\mathrm w.\mathrm r.\mathrm t.\;\;\;\boldsymbol\alpha,
\label{speech compressor}
\end{equation}
where $\boldsymbol\alpha$ is the trainable NN parameters of ${\mathfrak T}_{\mathrm{SC}}(\cdot)$.

The intermediate features, $\boldsymbol p$, are transmitted to the pre-trained large speech model on the edge server to extract semantic features $\boldsymbol f$ within a short period due to the exceptional computational prowess of the edge server. $\boldsymbol p$ is mapped to the symbols, $\boldsymbol x$, by the channel encoder. $\boldsymbol x$ can be expressed as
\begin{equation}
\boldsymbol x={\mathfrak T}_{\mathrm{CE}}(\boldsymbol f)\;\;\;\mathrm w.\mathrm r.\mathrm t.\;\;\;\boldsymbol\beta,
\label{channel encoder}
\end{equation}
where ${\mathfrak T}_{\mathrm{CE}}(\cdot)$ indicates the channel encoder and $\boldsymbol\beta$ is its trainable NN parameters.

The encoded symbols, $\boldsymbol x$, on the edge server are transmitted through the wireless fading channel. The received symbols, $\boldsymbol y$, on another edge server can be expressed as
\begin{equation}
\boldsymbol y=\boldsymbol h\ast\boldsymbol x+\boldsymbol n,
\label{channel}
\end{equation}
where $\boldsymbol h$ denotes the fading channel and $\boldsymbol n$ represents the additive white Gaussian noise (AWGN).

The channel decoder takes $\boldsymbol y$ as input and attains the estimated semantic features, $\widehat{\boldsymbol f}$, denoted as
\begin{equation}
\widehat{\boldsymbol f}={\mathfrak T}_{\mathrm{CD}}(\boldsymbol y)\;\;\;\mathrm w.\mathrm r.\mathrm t.\;\;\;\boldsymbol\gamma,
\label{channel decoder}
\end{equation}
where ${\mathfrak T}_{\mathrm{CD}}(\cdot)$ refers to the NN-based channel decoder.

Recovered $\widehat{\boldsymbol f}$ on the edge server is converted into the translated semantic information, $\widehat{\boldsymbol p}$, by the large speech model. Then, $\widehat{\boldsymbol p}$ is downloaded to the local device and retrieved by the speech predictor to generate the translated speech segment, $\widehat{\widetilde{\boldsymbol s}}_t$, expressed as
\begin{equation}
\widehat{\widetilde{\boldsymbol s}}_t={\mathfrak T}_{\mathrm{SP}}(\widehat{\boldsymbol p})\;\;\;\mathrm w.\mathrm r.\mathrm t.\;\;\;\boldsymbol\delta,
\label{speech predictor}
\end{equation}
where ${\mathfrak T}_{\mathrm{SP}}(\cdot)$ is the speech predictor and $\boldsymbol\delta$ represents its trainable NN parameters.

The translated speech segments are continuously provided to the receiver user to ensure seamless speech communication. It is noteworthy that the speech compressor and speech predictor are designed as lightweight NNs to alleviate the computational burden on the local device, which resolves challenge~\emph{C1}. The state-of-the-art large speech model is an ideal solution to address challenge~\emph{C2} because it is capable of extracting coherent semantic features and generating translated outputs across numerous languages. Additionally, the speech compressor and the speech predictor are pre-trained along with the large speech model without accounting for any communication issues while the channel encoder and the channel decoder are trained under specific channel conditions. The mean-squared error (MSE) is considered as the loss function to minimize the error between $\boldsymbol f$ and $\widehat{\boldsymbol f}$, modelled as
\begin{equation}
{\mathcal L}_{\mathrm{MSE}}(\boldsymbol f,\;\widehat{\boldsymbol f};\;\boldsymbol\beta,\;\boldsymbol\gamma)=\frac1L\sum_{l=1}^L{(f_l-{\widehat f}_l)}^2,
\label{loss function}
\end{equation}
where $L$ is the size of $\boldsymbol f$ and $\widehat{\boldsymbol f}$.

\subsection{Performance Metrics}
To assess the quality of the translated speech, the BLASER 2.0~\cite{dale-costa-jussa-2024-blaser} is adopted to measure the difference between the source and translated speech by returning calibrated and interpretable scores ranging from 1 to 5. BLASER 2.0 is tailored for multilingual speech translation and covers 57 spoken languages. Additionally, the average latency between two adjacent translated speech segments is calculated as a metric to evaluate the continuity of the translated speech and the overall communication latency, denoted as
\begin{equation}
\mathrm{AL}=\frac1{T-1}\sum_{t=1}^{T-1}{({\mathrm{TL}}_{{\widehat{\widetilde{\boldsymbol s}}}_{t+1}}-{\mathrm{TL}}_{{\widehat{\widetilde{\boldsymbol s}}}_t})},
\label{average lantency}
\end{equation}
where $T$ is the number of speech segments. ${\mathrm{TL}}_{{\widehat{\widetilde{\boldsymbol s}}}_{t+1}}$ and ${\mathrm{TL}}_{{\widehat{\widetilde{\boldsymbol s}}}_t}$ represent the transmission latency for $\widehat{\widetilde{\boldsymbol s}}_{t+1}$ and $\widehat{\widetilde{\boldsymbol s}}_t$, respectively.

\section{Large Model Empowered Semantic Communications for Speech Translation}
In this section, we introduce the LSSC-ST. Additionally, the proposed dynamic speech segmentation algorithm is presented.

\subsection{LSSC-ST}
\begin{figure*}[tbp]
\includegraphics[width=1.0\textwidth]{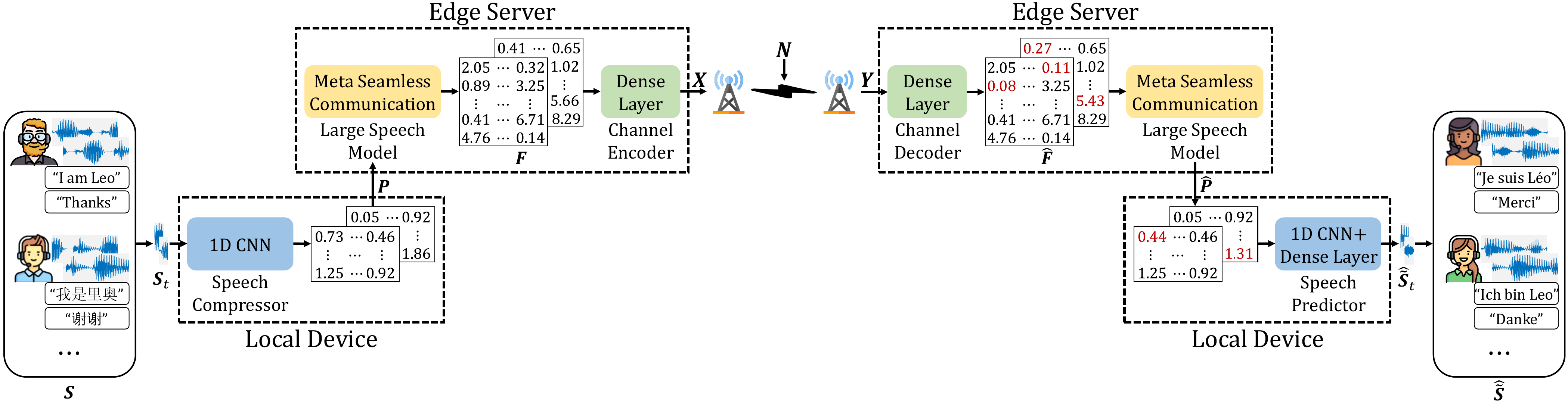}
\centering
\caption{Model structure of LSSC-ST for end-to-end streaming speech translation.}
\label{proposed system}
\end{figure*}

The proposed LSSC-ST is shown in Fig.~\ref{proposed system}. From the figure, the one-dimensional (1D) convolutional neural network (CNN)-based speech compressor condenses a batch of speech segments, ${\boldsymbol S}_t$, into the preliminary features, $\boldsymbol P$. The cutting-edge large speech model, Meta Seamless Communication~\cite{barrault2023seamless}, is deployed on the edge server and returns the learned semantic features, $\boldsymbol F$. Meta Seamless Communication supports many AI tasks, such as speech recognition and speech translation, across over 100 spoken languages. The dense layer-based channel encoder converts $\boldsymbol F$ into symbols $\boldsymbol X$ on the edge server before transmission over the wireless channel.

At the receiver, the obtained symbols, $\boldsymbol Y$, are passed through the dense layer-based channel decoder, and the output is the recovered semantic features, $\widehat{\boldsymbol F}$. The MSE loss is calculated after the channel decoder and backpropagated to the transmitter to update the trainable NN parameters of the channel encoder and decoder. The training algorithm is described in Algorithm~\ref{channel codecs training algorithm}. Next, the translated speech segments, $\widehat{\widetilde{\boldsymbol S}}_t$, are attained by feeding $\widehat{\boldsymbol P}$ into the 1D CNN and dense layer-based speech predictor. Finally, $\widehat{\widetilde{\boldsymbol S}}_t$ at each transmission is continuously concatenated to form a batch of translated speech, $\widehat{\boldsymbol S}$.
\begin{algorithm}[tbp]
\caption{Training algorithm of the channel encoder and decoder.}
\label{channel codecs training algorithm}
\textbf{Initialization:} Initialize trainable parameters $\boldsymbol\beta$ and $\boldsymbol\gamma$.
\begin{algorithmic}[1]
    \State \textbf{Input:} Batch of input speech $\boldsymbol S$, pre-trained speech compressor ${\mathfrak T}_{\mathrm{SC}}(\cdot)$ and large speech model, fading channel $\boldsymbol H$, Gaussian noise $\boldsymbol N$.
        \While{loss ${\mathcal L}_{\mathrm{MSE}}(\boldsymbol\beta,\;\boldsymbol\gamma)$ is not converged}
            \For{each batch of speech segments of ${\boldsymbol S}_t$} 
                \State ${\mathfrak T}_{\mathrm{SC}}({\boldsymbol S}_t)\rightarrow\boldsymbol P$.
                \State Upload $\boldsymbol P$ to the edge server.
                \State Extract $\boldsymbol F$ from $\boldsymbol P$.
                \State ${\mathfrak T}_{\mathrm{CE}}(\boldsymbol F)\rightarrow\boldsymbol X$.
                \State Transmit $\boldsymbol X$ over $\boldsymbol H$ and receive $\boldsymbol Y$ via (\ref{channel}).
                \State ${\mathfrak T}_{\mathrm{CD}}(\boldsymbol Y)\rightarrow\widehat{\boldsymbol F}$.
                \State Compute ${\mathcal L}_{\mathrm{MSE}}(\boldsymbol\beta,\;\boldsymbol\gamma)$.
                \State Update $\boldsymbol\beta$ and $\boldsymbol\gamma$.
            \EndFor
            \State \textbf{end for}
        \EndWhile
    \State \textbf{end while}
    \State \textbf{Output:} Trained ${\mathfrak T}_{\mathrm{CE}}(\cdot)$ and ${\mathfrak T}_{\mathrm{CD}}(\cdot)$.
\end{algorithmic}
\end{algorithm}

\subsection{Dynamic Speech Segmentation Algorithm}
In the LSSC-ST framework, the input speech is divided into multiple speech segments, each maintaining a constant duration. However, this rigid segmentation approach is not suitable in some extreme scenarios. For instance, when a speech segment contains considerable semantic information, the processes of generating $\boldsymbol F$ from $\boldsymbol P$ and obtaining $\widehat{\boldsymbol P}$ form $\widehat{\boldsymbol F}$ become highly time-consuming, thereby increasing the overall communication latency. To combat this, a dynamic speech segmentation algorithm is proposed to enable a more intelligent and adaptive mechanism for partitioning the input speech, which adjusts the current speech segment duration according to the semantic content of the previous segment. It is intuitive that the subsequent speech segment contains essential information if the amplitude of the speech samples in the current segment increases abruptly. Conversely, a decrease in amplitude suggests that the following segment may represent a pause in the speech flow. Additionally, when the amplitude of the current segment approaches zero, it implies that this segment carries no information.
\begin{figure}[tbp]
\centering
\begin{minipage}[c]{0.493\linewidth}
\centering
\includegraphics[width=1.0\textwidth]{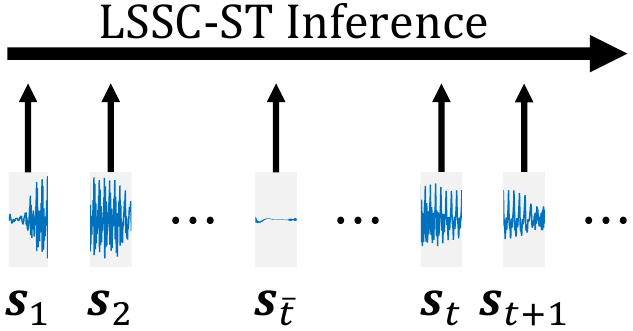}
\subcaption{}
\end{minipage}
\begin{minipage}[c]{0.493\linewidth}
\centering
\includegraphics[width=1.0\textwidth]{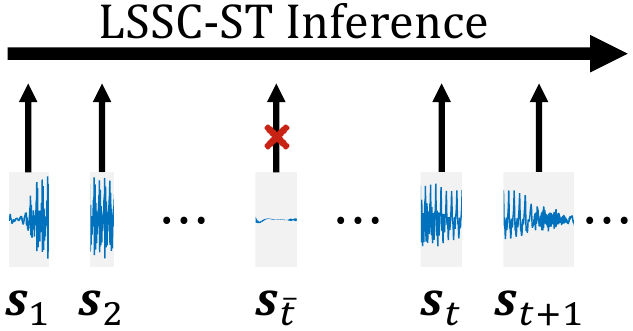}
\subcaption{}
\end{minipage}
\caption{(a) Fixed speech segmentation algorithm. (b) Dynamic speech segmentation algorithm.}
\label{dynamic algorithm}
\end{figure}

Inspired by this, an example of the dynamic speech segmentation algorithm is shown in Fig.~\ref{dynamic algorithm}. From the figure, the duration of the following segment is longer when the slope of the speech amplitude in the current segment exceeds zero, and it is shorter when the slope is smaller than zero. Silent segments are excluded from processing by the LSSC-ST during inference. Denote the duration of ${\boldsymbol s}_t$ to be $d_{{\boldsymbol s}_t}$, then $d_{{\boldsymbol s}_{t+1}}$ can be expressed as
\begin{equation}
d_{{\boldsymbol s}_{t+1}}=\left\{\begin{array}{l}\max\left[m,\;\left(1-pe^k\right)d_{{\boldsymbol s}_t}\right],\;\;\;\;\;\;\;\;\;\;\;\mathrm{if}\;k>0,\\\min\left[n,\;\left(1-q\ln\left(k+1\right)\right)d_{{\boldsymbol s}_t}\right],\;\;\mathrm{else},\end{array}\right.
\label{speech segment duration}
\end{equation}
where $k$ indicates the slope of the speech amplitude in the current segment, $p$ and $q$ are two hyperparameters, and $m$ and $n$ denote two thresholds that regulate the duration of speech segments, preventing them from becoming too long or too short, respectively.
\renewcommand\arraystretch{2.0} 
\begin{table}[tbp]
\footnotesize
\caption{Parameter settings of the proposed LSSC-ST.}
\label{parameter settings}
\centering
\begin{tabular}{|m{7.0em}<{\centering}|c|c|c|}
\hline
    \diagbox[dir=SW,width=8.5em,height=2.5em]{~}{~}            &   \textbf{Layer Name}   &   \textbf{Parameters}   &   \textbf{Activation}   \\
\hline
    \textbf{Speech Compressor}                                 &     2$\times$1D CNN     &     128, 64 channels    &           ReLU          \\
\cline{1-4}
    \textbf{Channel Encoder}                                   &     3$\times$Dense      &   3$\times$2048 units   &           None          \\
\cline{1-4}
    \textbf{Channel Decoder}                                   &     3$\times$Dense      &   3$\times$2048 units   &           ReLU          \\
\cline{1-4}
    \multirow{2}{7.0em}{\centering \textbf{Speech Predictor}}  &     2$\times$1D CNN     &  2$\times$1280 channels &           ReLU          \\
\cline{2-4}
                                                               &     1$\times$Dense      &         1 unit          &           None          \\
\cline{1-4}
    \textbf{Large Speech Model}                                &              \multicolumn{3}{c|}{Meta Seamless Communication}               \\
\hline
\end{tabular}
\end{table}

According to the dynamic speech segmentation, the testing algorithm of LSSC-ST for enabling streaming speech translation in semantic communications is illustrated in Algorithm~\ref{LSSC-ST testing algorithm}.

\section{Numerical Results}
The~\emph{FLEURS}~\cite{10023141} is adopted as the speech dataset to train and test LSSC-ST in the experiments. It covers speech utterances in 102 languages. Without loss of generality, we consider speech in English, Chinese, and French. The edge server and local devices are deployed within the campus local area network of Imperial College London. The edge server consists of six NVIDIA H100 GPUs.

The speech compressor consists of two CNNs. The channel encoder and decoder include three dense layers. Two CNNs and one dense layer are utilized in the speech predictor. The Meta Seamless Communication is invoked as the large speech model. Hyperparameters $p$ and $q$ are both set to 0.05. Thresholds $m$ and $n$ are 0.65 second and 0.85 second, respectively. The parameter settings of LSSC-ST are summarized in Table~\ref{parameter settings}.
\begin{algorithm}[tbp]
\caption{Testing algorithm of the LSSC-ST with dynamic speech segmentation mechanism.}
\label{LSSC-ST testing algorithm}
\begin{algorithmic}[1]
    \State \textbf{Input:} Batch of input speech $\boldsymbol S$, trained networks ${\mathfrak T}_{\mathrm{SC}}(\cdot)$, ${\mathfrak T}_{\mathrm{CE}}(\cdot)$, ${\mathfrak T}_{\mathrm{CD}}(\cdot)$, ${\mathfrak T}_{\mathrm{SP}}(\cdot)$, and large speech model, fading channel $\boldsymbol H$ from testing channel set $\boldsymbol{\mathcal H}$, a wide range of testing SNR regime.
    	\For{each testing SNR value}
            \For{each batch of speech segments of ${\boldsymbol S}_t$}
                \State Compute duration of ${\boldsymbol S}_t$ via (\ref{speech segment duration}).
                \If{${\boldsymbol S}_t$ is silent}
                    \State Break
                \Else
                    \State Generate Gaussian noise $\boldsymbol N$ under SNR value.
                    \State ${\mathfrak T}_{\mathrm{SC}}({\boldsymbol S}_t)\rightarrow\boldsymbol P$.
                    \State Upload $\boldsymbol P$ to the edge server.
                    \State Extract $\boldsymbol F$ from $\boldsymbol P$.
                    \State ${\mathfrak T}_{\mathrm{CE}}(\boldsymbol F)\rightarrow\boldsymbol X$.
                    \State Transmit $\boldsymbol X$ over $\boldsymbol H$ and receive $\boldsymbol Y$ via (\ref{channel}).
                    \State ${\mathfrak T}_{\mathrm{CD}}(\boldsymbol Y)\rightarrow\widehat{\boldsymbol F}$.
                    \State Attain $\widehat{\boldsymbol P}$ from $\widehat{\boldsymbol F}$.
                    \State Download $\widehat{\boldsymbol P}$ to the local device.
                    \State ${\mathfrak T}_{\mathrm{SP}}(\widehat{\boldsymbol P})\rightarrow\widehat{\widetilde{\boldsymbol S}}_t$.
                \EndIf           
            \EndFor
            \State \textbf{end for}
            \State Concatenate all $\widehat{\widetilde{\boldsymbol S}}_t$ to form $\widehat{\widetilde{\boldsymbol S}}$.
        \EndFor
        \State \textbf{end for}
	\State \textbf{Output:} Batch of translated speech, $\widehat{\widetilde{\boldsymbol S}}$.
\end{algorithmic}
\end{algorithm}

\begin{figure}[tbp]
\centering
\begin{minipage}[t]{0.493\linewidth}
\centering
\includegraphics[width=1.0\textwidth]{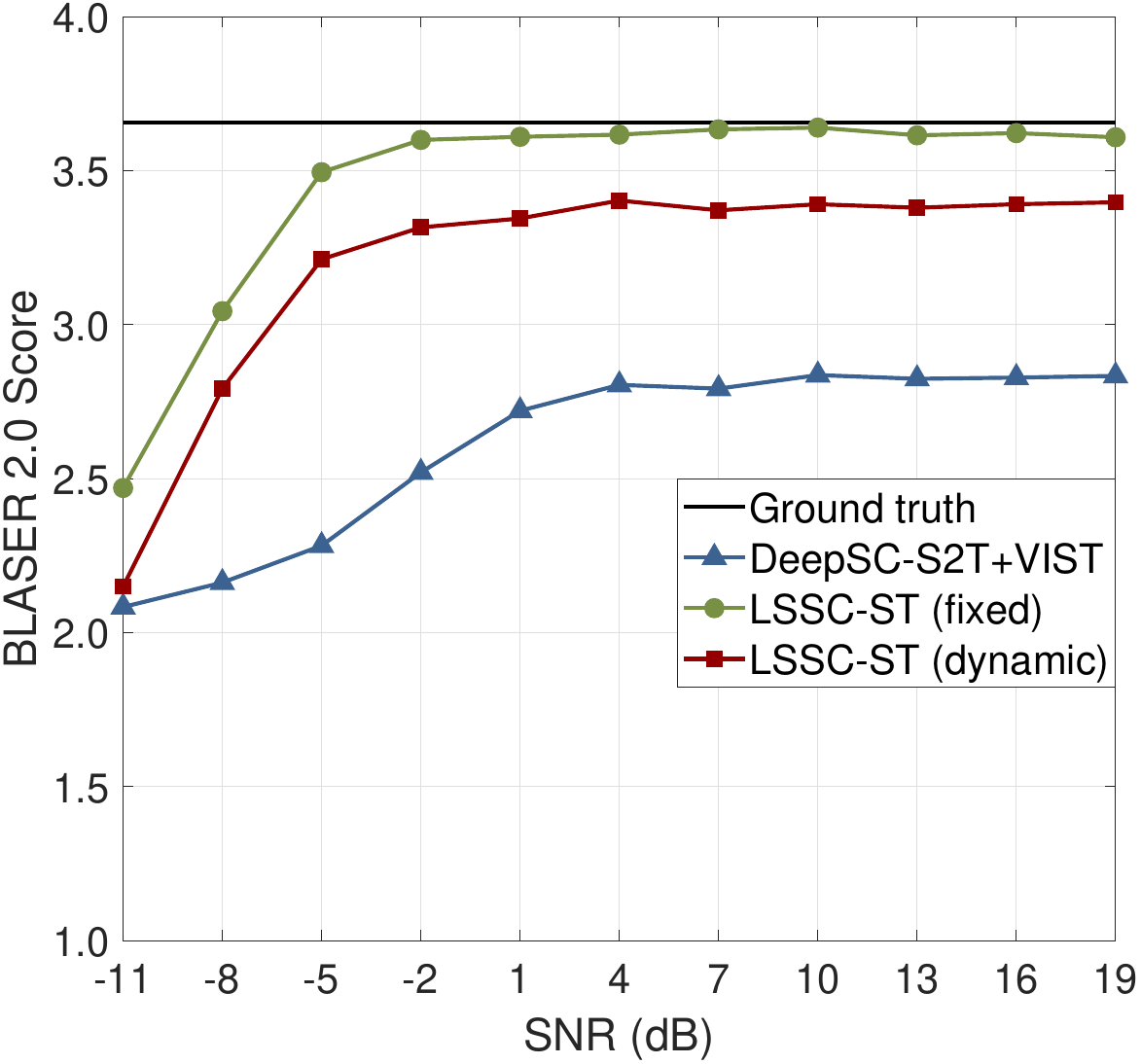}
\subcaption{AWGN channels}
\label{BLASER 2.0 result AWGN channels}
\end{minipage}
\begin{minipage}[t]{0.493\linewidth}
\centering
\includegraphics[width=1.0\textwidth]{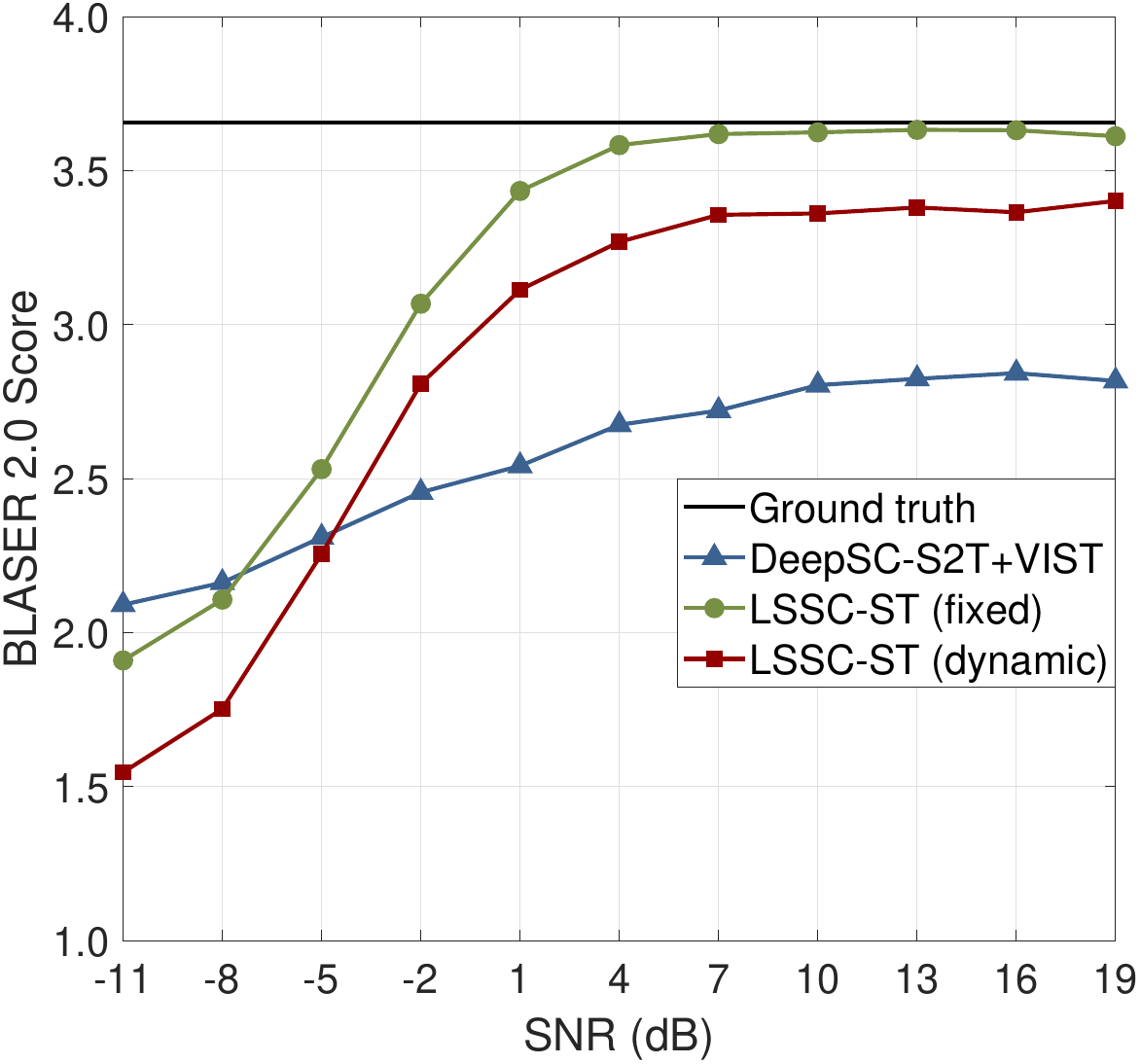}
\subcaption{Rayleigh channels}
\label{BLASER 2.0 result Rayleigh channels}
\end{minipage}
\caption{Simulation results of BLASER 2.0 scores.}
\label{BLASER 2.0 result}
\end{figure}

The BLASER 2.0 results are shown in Fig.~\ref{BLASER 2.0 result}, where the input language is English and the translated language is Chinese\footnote{More simulations results of different languages and the translated speech samples can be found at~\url{https://github.com/Zhenzi-Weng/LSSC-ST}.}. A benchmark is provided by cascading a semantic communication system for speech-to-text translation, DeepSC-S2T~\cite{weng2024robust}, and a cutting-edge text-to-speech pipeline, VIST~\cite{pmlr-v139-kim21f}. From the figure, the LSSC-ST outperforms the benchmark and attains BLASER 2.0 scores of over 3.0 in the high SNR regime, which verifies the effectiveness of the established edge-device collaborative semantic communication framework. Moreover, the LSSC-ST with the fixed speech segmentation approach provides superior quality of translated speech compared to the dynamic speech segmentation mechanism because the speech compressor and speech predictor are trained under the fixed-duration speech segments.
\renewcommand\arraystretch{2.0} 
\begin{table*}[tbp]
\footnotesize
\caption{Average latency of the translated speech segments predicted by different systems under Rayleigh channels.}
\label{average latency results}
\centering
\begin{tabular}{|c|c|c|c|c|}
\hline
    \diagbox[dir=SW,width=10.0em,height=2.5em]{~}{~}   &   \textbf{Ground Truth}   &   \textbf{DeepSC-S2T+VIST}   &   \textbf{LSSC-ST (Fixed)}   &   \textbf{LSSC-ST (Dynamic)}   \\
\hline
    eng-cmn                                            &           0.33            &            10.67             &             0.49             &         \textbf{0.42}          \\
\hline
    eng-fra                                            &           0.34            &      \textbf{\texttimes}     &             0.48             &         \textbf{0.40}          \\
\hline
    cmn-eng                                            &           0.42            &      \textbf{\texttimes}     &             0.57             &         \textbf{0.46}          \\
\hline
    cmn-fra                                            &           0.33            &      \textbf{\texttimes}     &             0.45             &         \textbf{0.37}          \\
\hline
    fra-cmn                                            &           0.29            &      \textbf{\texttimes}     &             0.45             &         \textbf{0.38}          \\
\hline
    fra-eng                                            &           0.34            &      \textbf{\texttimes}     &             0.49             &         \textbf{0.41}          \\
\hline
\end{tabular}
\vspace{0.2em}
\caption*{{\footnotesize eng is English, cmn is Mandarin, and fra is French. eng-cmn refers to speech translation from English to Mandarin.}}
\vspace{-1.0em}
\caption*{{\footnotesize All values in the table are in seconds.}}
\end{table*}

The average latency results of different systems are presented in Table~\ref{average latency results}. From the table, the benchmark has a latency of over 10 seconds due to the requirement for the entire input speech, and it merely supports speech translation from English to Chinese. The LSSC-ST with the dynamic speech segmentation algorithm achieves an average latency of around 0.4 seconds across all tested languages, reducing transmission latency by 14.3$\%$ to 19.3$\%$ compared to fixed speech segmentation scenarios. Therefore, the LSSC-ST, with the proposed dynamic speech segmentation algorithm, offers a promising solution for enabling low-latency multilingual speech translation in semantic communications.

\section{Conclusions}
In this paper, we developed a large model-empowered semantic communication system to support streaming speech transmission, named LSSC-ST. Particularly, the semantic extraction and channel coding modules are offloaded to the edge server to mitigate the computational demands on the local device. The large speech model is leveraged to break the language constraint of the input speech, generating unified semantic features and supporting multilingual speech translation tasks. Moreover, a novel dynamic speech segmentation algorithm reduces the end-to-end transmission latency by adaptively adjusting the speech segment duration.

\bibliographystyle{IEEEtran}
\bibliography{reference.bib}

\end{document}